# Study of depth profile of hydrogen in hydrogenated diamond like carbon thin film using ion beam analysis techniques

J Datta[a], H S Biswas[b], P Rao[c], G L N Reddy[c], S Kumar[c], N R Ray[d,*], D P Chowdhury[a], A V R Reddy[e]

[a]Analytical Chemistry Division, BARC, Variable Energy Cyclotron Centre, 1/AF, Bidhan Nagar, Kolkata 700064, India

[b]Dept. of Chemistry, Surendranath College, Kolkata-700009, India

[c]National Centre for Compositional Characterization of Materials, BARC, ECIL Post, Hyderabad 500062, India

[d]Surface Physics Material Science Division, Saha Institute of Nuclear Physics, 1/AF, Bidhan Nagar, Kolkata 700064, India

[e]Analytical Chemistry Division, Bhabha Atomic Research Centre, Trombay, Mumbai 400085, India

*Corresponding author: niharranjan.ray@saha.ac.in

**Abstract.** The Hydrogenated Diamond Like Carbon (HDLC) thin films are deposited on Silicon substrate at room temperature using asymmetric capacitively coupled RF plasma with varying flow rates of methane. These films are undergone annealing at high vacuum ($\sim 10^{-7}$ torr) and high temperature (750 and 1050 $^0$C) furnace. The as-prepared and annealed HDLC films have been depth profiled for hydrogen using the resonance at 6.44 MeV in $^1$H($^{19}$F,$\alpha\gamma$)$^{16}$O nuclear reaction. The as prepared films exhibit non-uniform depth distribution of hydrogen: it decreases with depth. Annealing in vacuum brings about is a significant desorption of hydrogen from the films. Loss of hydrogen, albeit in much lower proportions, is also induced by the bombarding beam. The films also experience a mild loss of carbon, as shown by proton backscattering spectrometry, during high vacuum annealing. The depth profiles of hydrogen in the annealed films are indicative of the prevalence of graphitic carbon near film –substrate interface.

## 1. Introduction

In recent years, diamond like carbon (DLC) films has attracted a great deal of research interest as they have enormous possibility for technological applications [1, 2]. Depending upon atomic structure of DLC films, the behavior of the film is diamond like. For example, opaque samples with hardness one-fifth of that of diamond and the transparent ones nearly as dense and hard as diamond [3] have considered as DLC films. The films with 20% to 40% of hydrogen content are commonly known as hydrogenated DLC (HDLC) films. In general, the physical and chemical properties of carbon materials, having sp$^3$ (diamond like) and sp$^2$ (graphite like) bonds in different ratio in the carbon matrix, may be different compared to pure graphite or diamond. We have demonstrated earlier [4, 5] that HDLC films, deposited onto Si (100) substrate by reactive gas-plasma process, are composed of an ordered hexagonal structure of carbon atoms with lattice parameters a= 2.62 Å and c=6.752 Å different to those present in a hexagonal graphite structure. Further structural investigation by confocal micro Raman spectroscopy [5, 6] shows that coherency of sp$^3$ C-H and sp$^2$ C=C carbons in the HDLC can produce a continuous nonporous thin film (thickness ≈ 168 nm) having atomically smooth surface. Recently, the signature of interlayer disorder region in the HDLC film has been observed when Raman spectra of as-prepared HDLC and annealed HDLC samples are compared [5, 6]. Due to existence of interlayer disorder, the 3D crystallinity of the HDLC should be lost, while only the 2D crystallinity should be preserved [7]. The Raman spectra do not tell us about why and how the interlayer disorder region is created in the HDLC film. The main effect of hydrogen in HDLC films is to modify its C-C network at different depth of the film. How these modifications occur at various depth of the HDLC film during its synthesis onto Si(100) substrate cannot be studied by Raman spectroscopy or X-ray photoelectron (XP) spectroscopy method, because (i) the skin depth of the 488 nm excitation source, which we have used in Raman measurements, is ~ 6 μm and using this technique we get information for the whole material rather than at different depth of the thin film having thickness ~ 168 nm (ii) XP spectroscopy [8, 9] is a surface sensitive probe in our case, because here the skin depth for X-rays is ~5 nm. These earlier works [5-7] motivate us to explore the distribution of hydrogen in the film along its depth by ion beam analysis techniques, viz., nuclear reaction analysis (NRA), Rutherford backscattering (RBS) techniques, in order to know how hydrogen modify C-C network in the HDLC film onto Si(100) substrate.

## 2. Experimental

The HDLC thin films are deposited on mirror-polished Si (100) substrate at room temperature using asymmetric capacitively coupled RF 13.56 MHz plasma system. The depositions are made systematically as follows: a pretreatment of the bare mirror-polished Si(100) substrate has been done for 15 min using pure hydrogen plasma at pressure of 0.2 mbar and dc self-bias of - 200 V. The deposition has been made for 30 min at pressure of 0.7 mbar keeping the flow rate of helium (He) at 1500 SCCM (SCCM denotes cubic centimeter per minute at STP), hydrogen (H$_2$) at 500 SCCM, and varying the flow rate of methane (CH$_4$).

Four samples for which the CH$_4$ flow rates are 20, 30, 40 and 70 SCCM, thus grown with varying H$_2$ to CH$_4$ ratio during deposition at room temperature (RT), will be represented as samples A, B, C and D respectively, in the rest of the article. The samples A-D, annealed at high temperatures 750 $^0$C and 1050 $^0$C in high vacuum (~ 1×10$^{-7}$ torr) furnace, in order to study removal of hydrogen from the samples, will be represented as annealed (750 $^0$C / 1050 $^0$C) samples A-D in the rest of the article.

Nuclear reaction analysis (NRA) [10-11], a non-destructive nuclear method for depth profiling, has been applied for the quantitative determination of hydrogen at different depths of the as-prepared and annealed samples A-D, using the resonance at 6.44 MeV for the reaction $^1$H($^{19}$F,αγ)$^{16}$O in the present experiment. The Rutherford backscattering (RBS) [12-14] on the as-prepared and annealed samples A-D, with 1.0 MeV proton beam is carried out to measure the thickness of HDLC films. Both NRA and RBS measurements are carried out by a 3 MV Tandetron accelerator at the surface and profile measurement laboratory of the National Center for Compositional Characterization of Materials (NCCCM), Hyderabad, India.

*2.1. NRA measurements*

Depth profiling of hydrogen in the samples A-D are accomplished by bombarding the sample at normal incidence with a well-collimated (dia. 2 mm, current 3 nA) tripositive fluorine ion ($^{19}$F$^{3+}$) beam in the scattering chamber under vacuum 10$^{-6}$ mbar. The characteristic gamma rays for the reaction $^1$H($^{19}$F,αγ)$^{16}$O are monitored using a bismuth germanate (BGO) semi-conductor detector having efficiency ~10% and placed at a distance of 2 cm behind the sample along the direction of the incident beam. The characteristic gamma rays from the produced isotope $^{16}$O are 6.1, 6.9 and 7.1 MeV. A guard ring, with a bias of - 900 V is positioned in front of the sample to suppress secondary electrons. The beam current incident on the sample is measured by Faraday cup arrangement [15]. The γ-ray yield of the reaction was obtained from the integrated counts between 4.8 and 7.1 MeV energy window of the PC-based multichannel analyzer, and this is found to give significant counts in the present experimental set.

The measurements for depth profiling in the sample and standard (viz., mylar) [16-18] are performed in the following steps: (i) γ−yield is measured at off-resonance region (beam energy below 6.44 MeV) to have the contribution of the background for each sample (ii) energy of the incident ion beam is increased in steps of 20 keV/40 keV above the resonance energy 6.44 MeV till the integrated counts in the region of interest are equal or less than that of the off-resonance counts. In the above two steps, the typical beam currents for both sample and standard are of the order of 3 and 2 nA over a ~ 2 mm diameter spot. Two sets of measurement (in steps of 20 and 40 keV) are carried out for each sample in the different regions to get the reproducibility of the results. The desorption phenomenonon is studied for our specimen and standard which are treated at the same conditions of the beam and the integrated counts in the multichannel analyzer are recorded in every 250 nC, until total charge in the Faraday cup is 5000 nC.

*2.2. RBS measurements*

RBS measurements on HDLC thin films (as prepared and annealed) are carried out in the scattering chamber under vacuum ~$5\times10^{-6}$ mbar by with 1.0 MeV proton beam from the same Tandetron accelerator. The backscattered particles are detected by a silicon surface barrier (SSB) detector at an angle of $170^0$.

*2.3. Methodology and Technical Details*

*2.3.1. Expression for Hydrogen content determination using NRA technique.* The atomic fraction of hydrogen in the HDLC film (considering as a binary film, $C_xH_y$) is estimated by the following equation [19]

$$f^H_{HDLC} = \frac{f^H_{Std} \cdot \varepsilon^C_{HDLC}}{Y\varepsilon_{std} + f^H_{Std} \cdot (\varepsilon^C_{HDLC} - \varepsilon^H_{HDLC})} \quad (1)$$

where $Y$ is the ratio of gamma yields for standard and sample, $f$ is the atomic fraction of hydrogen in the film material given in subscript, $\varepsilon_{Std}$ and $\varepsilon^i$ are the stopping cross section for the standard and the $i^{th}$ element having in the superscript. The ratio of yields, $Y$ is obtained from the γ-ray yields in the NRA measurements. The stopping cross section and atomic fraction of hydrogen for the standard Mylar $\varepsilon_{std}$ $1.841\times10^2$ eV/($10^{15}$ at./cm$^2$) and $f^H_{std}$ 0.3636 respectively, are known. The stopping power data in this work is calculated using SRIM-2008.04 [20].

The depth $\chi_R$ is related to the incident beam energy $E_i$ by the equation

$$\chi_R = \frac{E_i - E_R}{\varepsilon} \quad (2)$$

where $E_R$ is the resonance energy and $\varepsilon$ is the stopping power. The depth in at./cm$^2$ is converted into linear dimension using density of the film of known film thickness determined by other methods.

The stopping power of a multi-elemental film using Bragg's law of linear additively can be written as

$$\varepsilon = \sum_{i=1}^{N} f_i \cdot \varepsilon_i \quad (3)$$

where $f_i$ and $\varepsilon_i$ are the atomic fraction and the atomic stopping cross section of the i[th] constituent, respectively and N is the total no. of constituents in the film.

*2.3.2. Expression for thickness determination using RBS technique.* The energy loss of a 1.0 MeV proton beam in a thin film of thickness t, related to the stopping factor of the material is given below [13-14].

$$t = \frac{\Delta E}{[\varepsilon_i]} \tag{5}$$

$$[\varepsilon_i] = \frac{K_i}{\cos\theta_1}\left[\frac{dE}{dX}\right]_{in} + \frac{1}{\cos\theta_2}\left[\frac{dE}{dX}\right]_{out} \tag{6}$$

where t is the thickness, $\Delta E$ is the FWHM of the peak, $\varepsilon_i$ is the stopping cross section of the i[th] element

## 3. Results and Discussions

*3.1. NRA measurements*

Figure 1 shows typical γ-ray yield (due to nuclear reaction $^1H(^{19}F,\alpha\gamma)^{16}O$), as a function of (i) energy difference between energy of incident $^{19}F$ ion beam ($E_{in}$) and the resonance energy ($E_R$) as recorded in a PC-based multichannel analyzer, (ii) hydrogen concentration at different depth of the film, estimated using stopping power value of HDLC. For $E_{in} < E_R$, the γ-ray yield corresponds to 'background yield' and for $E_{in} = E_R$ the γ-ray yield corresponds to the presence of hydrogen atoms onto the surface of HDLC sample. With the increase of energy of the beam penetrating into the sample, the γ-ray yield starts decreasing from the surface of the target, as evident in Fig.1. The typical behavior of declining of γ-ray yield at different depth of as prepared and annealed HDLC sample are observed to be similar.

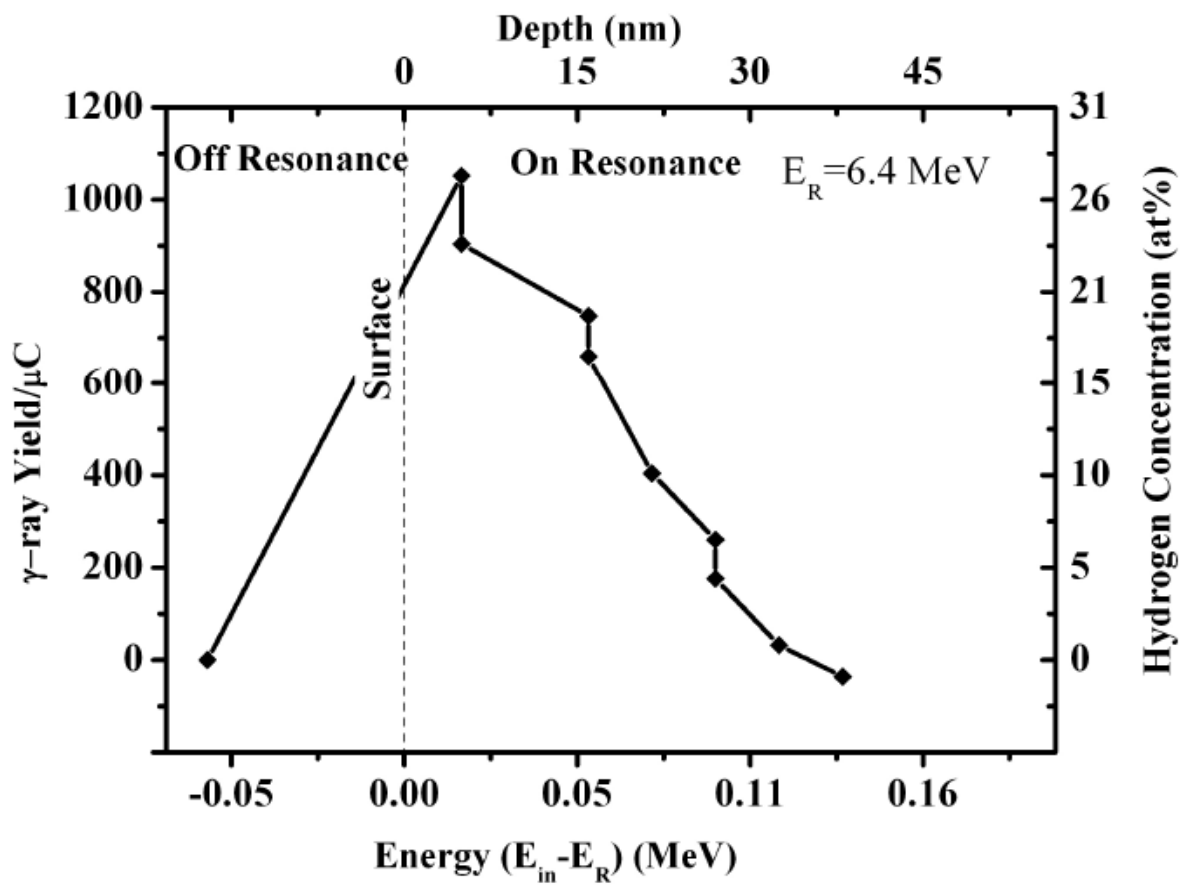

**Figure 1. The typical NRA profile of as prepared HDLC sample-A is shown. Both raw data of 6.1 MeV prompt γ-ray yields per μC of charge Vs. difference of incident beam energy and resonance energy (6.4 MeV), and corresponding hydrogen concentration (at.%) Vs. depth (nm) into HDLC sample are shown above.**

The variation of yield with depth is faster in as-prepared sample than that for annealed one, as shown in figure 2. The as-prepared samples (A to D) when annealed at different temperatures 750 $^0$C and 1050 $^0$C in high vacuum, show that the yields, and hence the hydrogen content of the annealed films, in all the cases, are lower than that of the as-prepared samples. A typical result for the as-prepared sample B is shown in figure 2. As expected, the hydrogen content should decrease with the increase of the annealing temperature. Therefore, the results in figure 2 clearly indicate significant loss of bonded hydrogen atoms from the HDLC sample due to annealing effect. Since the decades, it has been well known that the heating of amorphous hydrogenated carbon above 300 $^0$C will produce the exo-diffusion of hydrogen molecules through the recombination of small sized hydrogen atoms within the bulk of the materials. Such formation of $H_2$ molecules is favorable due to the higher bond energy of H-H than that of the C-H bond. Therefore, annealing leads to the transformation of $sp^3$C-H sites into $sp^2$ sites. Consequently, there is the reduction of the photoluminescence background, validated in our earlier results [5]. It is also to be noted from the figure 2 that the maximum depth at which the yields of hydrogen content are more or less equal to the background counts depends upon the annealing temperature; these depths are ~ 83 nm for the as prepared sample, ~68 nm for annealing temperature 750 $^0$C and ~ 59 nm for annealing temperature 1050 $^0$C. It is well established that the diamond is metastable and graphite is thermodynamically stable, and therefore, diamond and/ nano-crystalline Diamond (NCD) are converted into graphite when heated at elevated temperature ca 1300 $^0$C [21-23]. Therefore, our results showing decreasing of depth of hydrogen content in the HDLC sample with increasing annealing temperature, may be attributed to conversion of sp$^3$ C-H (diamond like) to sp$^2$ C=C (graphite like) structure in the HDLC sample.

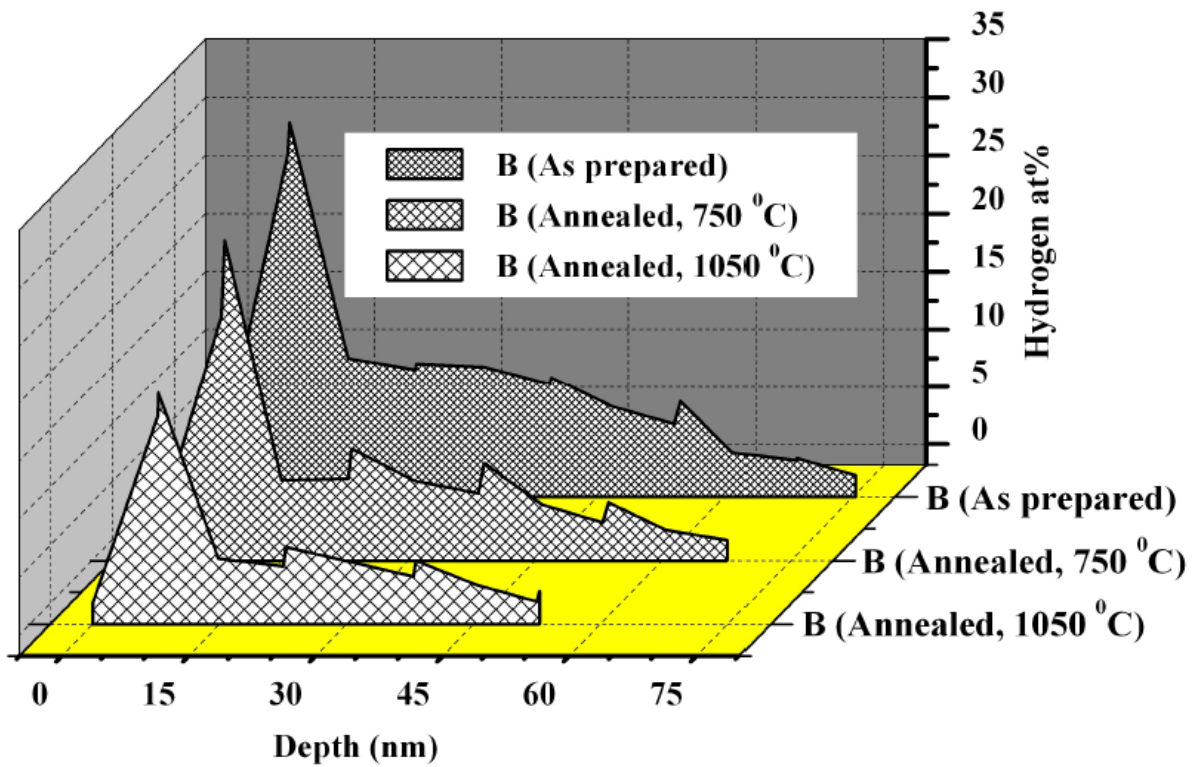

**Figure 2. NRA measurements on HDLC sample-B thin films onto Si backing materials (30 SCCM, both as prepared and annealed at 750 and 1050 $^0$C) were carried out with $^{19}$F$^{3+}$ ion beam at beam current of 3 nA. For each of the above measurements fresh samples were annealed.**

The variation of hydrogen content with the depth of the as-prepared samples (A-D) is shown in figure 3. The hydrogen concentration is maximum just inside the surface which is approximately within 10 nm from the surface in most of the samples. This result may indicate that the unsaturated bonds of carbon are being terminated by hydrogen, thus leaving almost no unsaturated bonds in the samples at the end of their deposition. From the figure 3, it is evident that at a given depth within the sample, the hydrogen content increases with the increase of flow rate of methane at a constant flow rate of hydrogen (500 sccm) during deposition of samples (A-D). This result is in accordance with our earlier results as reported Singha *et al* (2006) [24]. The continuing decrease of the hydrogen with depth of the samples (A-D) indicates that the films having higher graphitic nature towards the interface between the HDLC film and Si(100) substrate. This observation could be explained considering the fact that, in a bias enhanced nucleation of diamond on Si, there is always one intermediate buffer layer of SiC or graphite as shown in a model of diamond/graphite/Si(100) interface [25-26]. This model provides the minimal lattice mismatch and minimal interfacial energy.

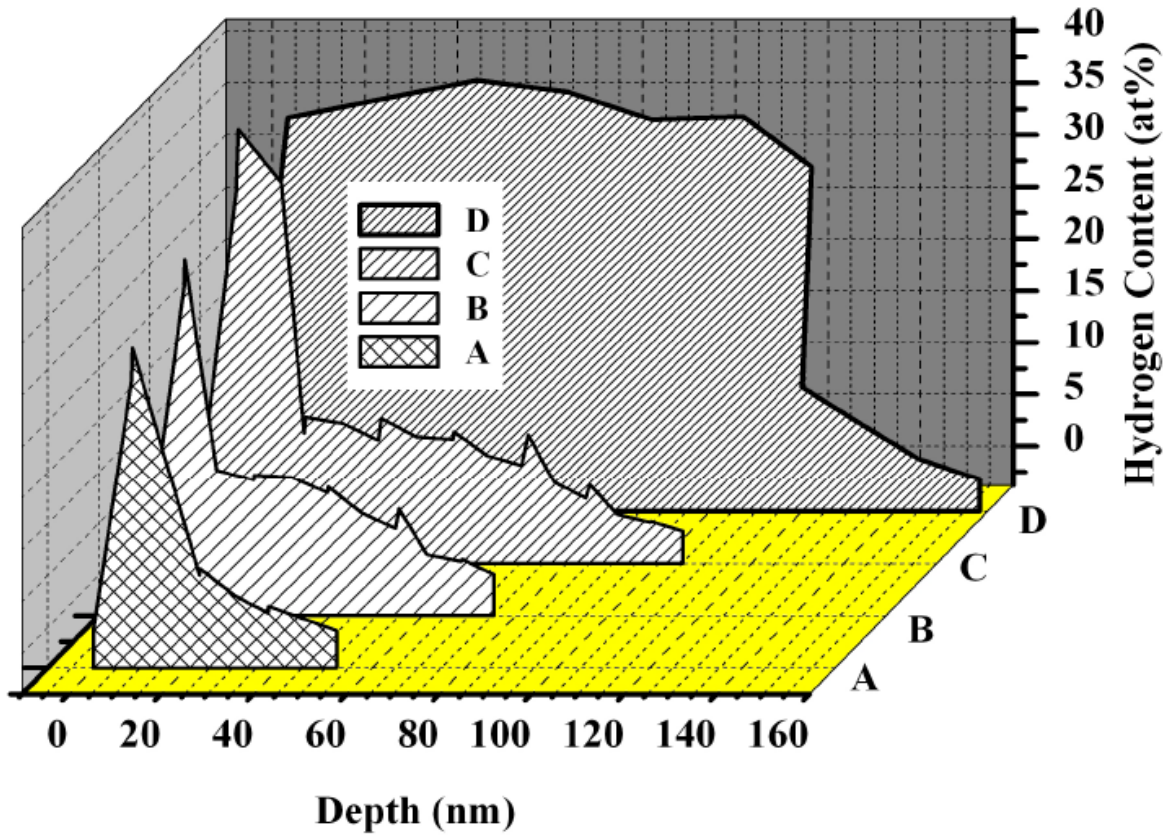

**Figure 3. Hydrogen concentration distribution observed in NRA measurements for as prepared HDLC samples (A-D) deposited in different flow rate (SCCM-standard cubic centimeter per minute) with $^{19}F^{3+}$ ion beam at beam current of 3 nA.**

The maximum depth of hydrogen content in the HDLC film depends upon flow rate of methane (figure 3) and for a given as-prepared HDLC film the maximum depth of hydrogen content decreases with increasing annealing temperature (figure 2). Therefore, the direction of flow out of hydrogen during annealing of HDLC sample is from the interface towards the surface. E. Vainonen-Ahlgren *et al* (1997) [27] and J P Thomas *et al* [29] have studied the migration of hydrogen in DLC films by NRA, RBS, and SIMS techniques and found migration of hydrogen towards the interface between the film and substrate and also release of hydrogen from the surface region. On contrary to this observation, the migration of hydrogen towards surface from the interface in our HDLC sample seems to be plausible due to the fact that during the annealing process, the heat is transferred from the sample holder to the backing silicon material and then into the sample; hence, the interface of the sample and the Si is at higher temperature than the surface of the HDLC samples. Therefore, there might be a chance of little rearrangement of hydrogen bonds through the process of dehydrogenation and hydrogenation [30] within the sample. This effect seems to be is neglected in the earlier works [5-6].

In some hydrogenous carbon materials loss of hydrogen under ion irradiation is reported earlier [31-32]. Due care, therefore, is to be taken while carried out the hydrogen determination. In such cases, the concentration of hydrogen should be measured as a function of ion dose and the initial hydrogen content can be obtained by extrapolation to the zero dose condition. This result is shown in figure 4. The yields at zero dose condition are also shown along the dotted line for the as-prepared and annealed samples. These values were obtained from the polynomial fitting of the γ-ray yields vs. charge density. The stability of the standard samples under ion beam irradiation has already been studied [16]. This can be minimized by using the lower mass of the projectiles, which don't transfer large electronic energy as excitation energy in the sample. Nearly constant integrated counts (as shown in figure 4) for every 250 nC charge, were observed for the standards ($Si_3N_4$) for total 20 number of times exposure at primary beam energy of ~ 6.5 MeV. In contrary, the counts were slowly decreased for annealed samples and rapidly for the as-prepared samples as shown in figure 4. This could be explained if we assume that the hydrogen is present in the film as bonded by chemical forces and also as non-bonded due to physical forces or weak Vander Wall forces. Therefore, the weakly bonded hydrogen readily desorbs from the thin films on heating. This desorption is maximum for the as-prepared films compared to annealed films.

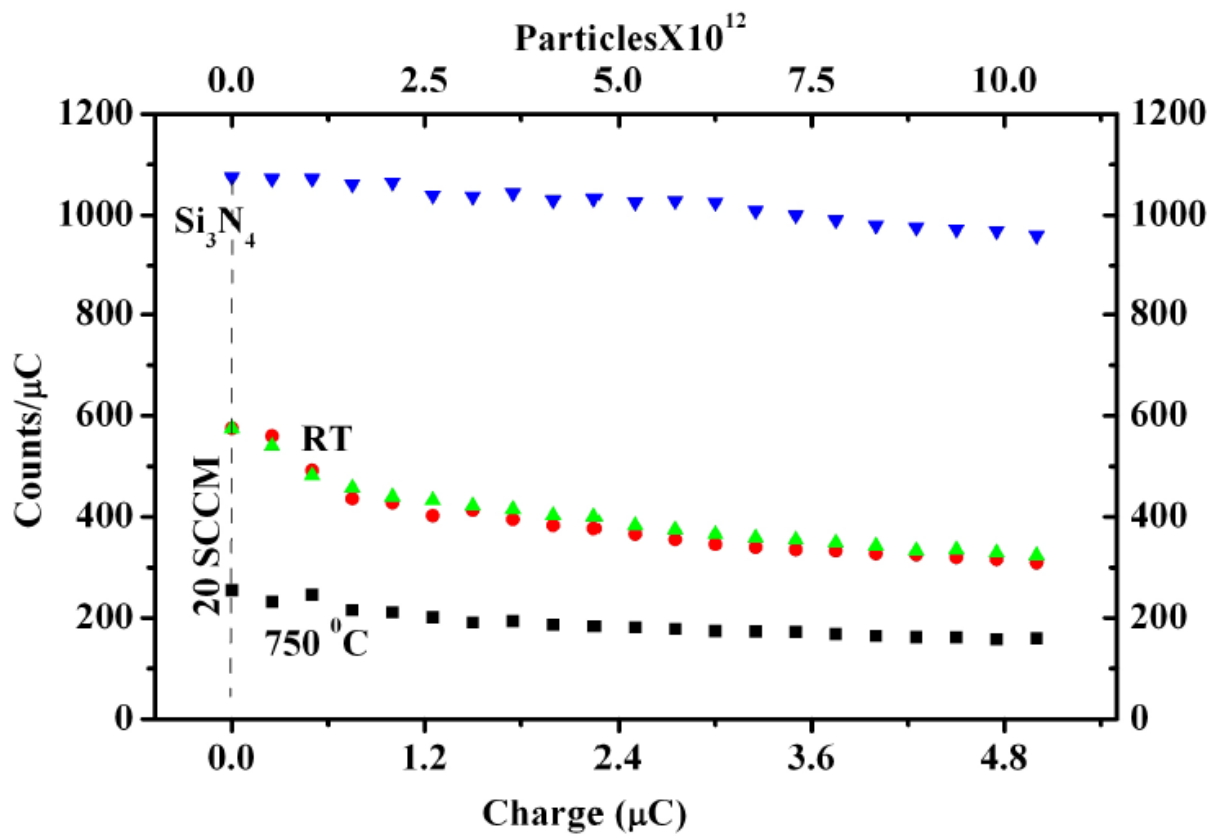

**Figure 4. The γ-ray counts/µC vs. charge (µC) (Bottom) or fluence of $^{19}F^{3+}$ ion beam (Top) during hydrogen depth profiling for as prepared (at two different regions of 5 mm apart) and annealed at 750 $^0$C of HDLC samples-A at beam current of 3 nA and also Si$_3$N$_4$ standard at beam current of 3 nA observed in NRA technique.**

*3.2. RBS measurements*

The backscattered spectra for as prepared and annealed samples are recorded using 1.0 MeV proton beam and are shown in figure 5. The higher channel number (i.e., higher energy) side of the RBS spectrum is due to the backing silicon materials (thickness ~2000 nm). The sharp peak is observed over the broad plateau of silicon is due to the carbon. It is clear from the inset picture where there is no peak for the virgin silicon sample but sharp peaks for carbon onto silicon. The sharp peaks of carbon in the inset of figure 5, show increase of counts with the increase of flow rates of methane in the as-prepared samples. Therefore, the thickness of the film increases with flow rates of methane.

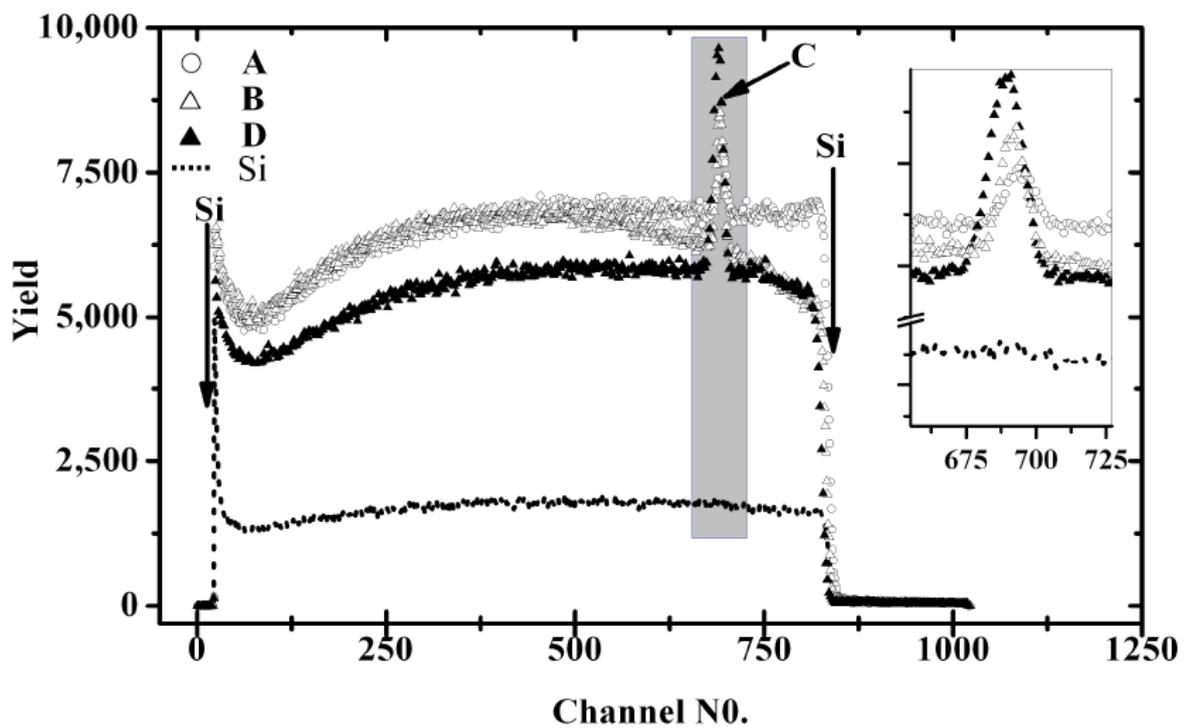

**Figure 5. RBS measurements on as prepared HDLC thin films onto Si backing materials prepared at different flow of methane were carried out with 1.0 MeV proton beam. The backscattered particles were detected by a silicon surface barrier (SSB) detector at an angle of 170$^0$.**

The RBS spectra for the as prepared and annealed HDLC samples are measured. It is observed that the as prepared films have higher peak yields and higher FWHM compared to that of annealed one as shown in figure 6. This result indicates thickness of as-prepared samples is larger than that for annealed one. Thickness of annealed films at 750 and 1050 $^0$C are nearly equal.

Kazunori *et al* (1999) [33] and Wang *et al* (1997) [34] have studied the erosion behavior of deuterated soft amorphous carbon thin film by heat treatment in air and vacuum, using elastic recoil detection (ERD) and proton enhanced scattering (PES) techniques. When the films are heated in air above 550K for 1 h, the film thickness, and the areal densities of carbon and deuterium both decrease. Moreover oxygen is incorporated into the films. But when annealed in vacuum, erosion starts at above 600K and decrease in thickness along with the decrease of all other components. Thermal desorption spectroscopy of the soft films reveals the evolution of hydrocarbon. They have also studied the erosion behavior of hard a-C:H thin films when annealed under vacuum, release predominantly hydrogen molecule to give semi-hard a-C:H films with very low hydrogen content. This implies that the soft films sublimate and hard film desorbs in vacuum during annealing. We have also observed the similar desorption and decrease in at% of hydrogen and that in thickness for HDLC thin film, when treated under vacuum at high temperature. The diagnosis by NRA measures the content of hydrogen whereas RBS techniques measures content of carbon. Therefore, the decrease of HDLC film thickness due to annealing (figure 6) should be caused by removal of carbon and hydrogen both.

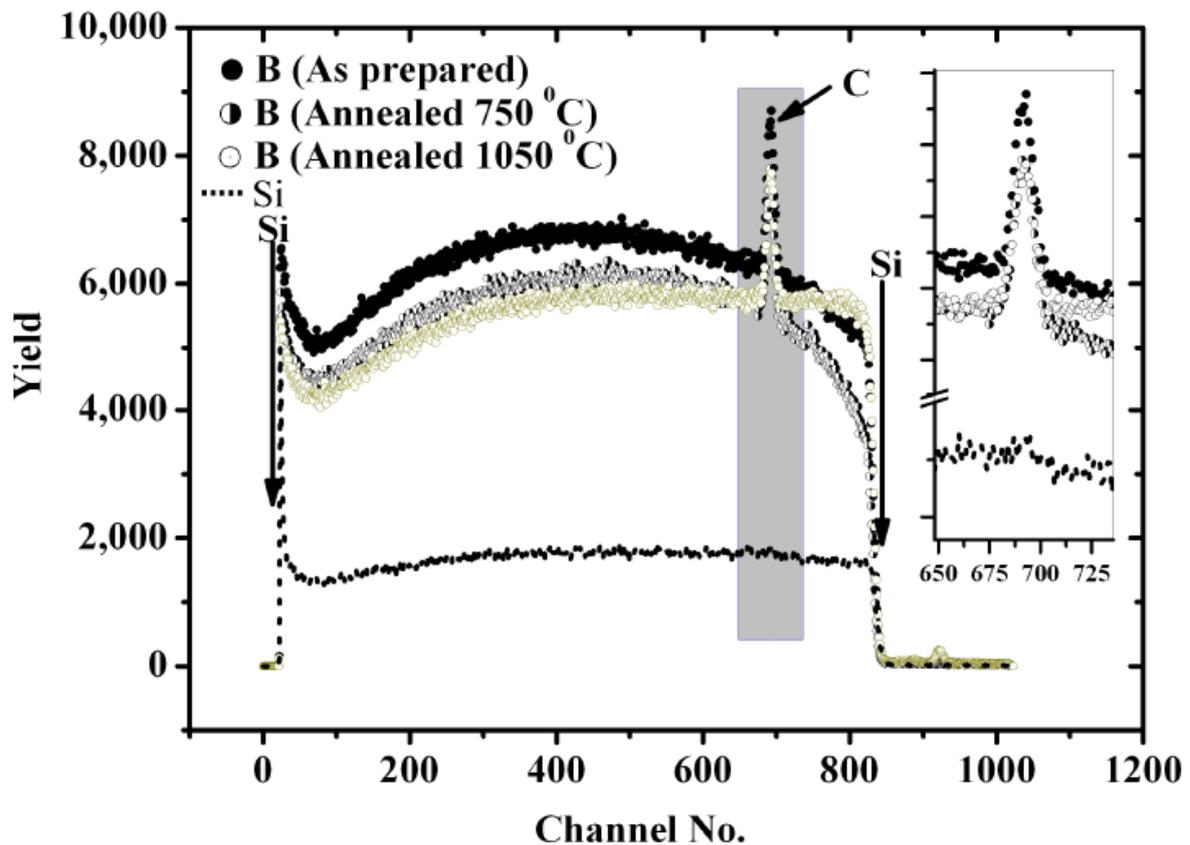

**Figure 6. RBS measurements on HDLC thin films B onto Si backing materials (30 SCCM, both as prepared and annealed at 750 and 1050 $^0$C) were carried out with 1.0 MeV proton beam from**

Tandetron accelerator. The backscattered particles were detected by a silicon surface barrier (SSB) detector at an angle of $170^0$. For each of the above measurements fresh samples were annealed.

*3.3. Estimation of hydrogen content and thickness of HDLC sample*

The atomic fraction of Hydrogen in the as prepared HDLC film is estimated using equation (1) based upon NRA measurements. The results are given in Table 1.

**Table 1. Atomic fraction of hydrogen in the HDLC films (A-D) as a function of flow rate of methane**

| Sample | Flow rate of methane (SCCM) | Hydrogen content At.% |
|--------|------------------------------|----------------------|
| A      | 20                           | 27.1                 |
| B      | 30                           | 30.5                 |
| C      | 40                           | 32.9                 |
| D      | 70                           | 34.1                 |

The thickness of the as prepared HDLC film is measured by NRA using equation (2) and by RBS using equation (5). These results are given in Table 2.

**Table 2. Comparison of depth of hydrogen by NRA and thickness of the as prepared films (A-D) by RBS technique**

| Sample | Flow rate of $CH_4$ (SCCM) | Depth of hydrogen in the film by NRA (nm) | Thickness of the film by RBS (nm) |
|--------|----------------------------|-------------------------------------------|-----------------------------------|
| A      | 20                         | 52                                        | 124                               |
| B      | 30                         | 83                                        | 130                               |
| C      | 40                         | 98                                        | 134                               |
| D      | 70                         | 158                                       | 144                               |

Thickness of HDLC film by NRA is based upon maximum depth of hydrogen availability whereas that by RBS is based upon maximum depth of carbon availability in the film. Since toward interface between substrate and HDLC film, graphitic ($sp^2$ C=C) character is predominant than diamond ($sp^3$ C-H) character, it is obvious that the observed thickness by NRA smaller than that by RBS (table 2) is reasonable.

**Table 3. Effect of annealing on the thickness of the HDLC thin films**

| Sample | Flow rate of $CH_4$ (SCCM) | Depth of hydrogen in the film by NRA (nm) | | |
|---|---|---|---|---|
| | | RT ($^0$C) | 750 ($^0$C) | 1050 ($^0$C) |
| A | 20 | 52 | 48 | 36 |
| B | 30 | 83 | 68 | 59 |
| C | 40 | 98 | 82 | 73 |
| D | 70 | 158 | 150 | 145 |

The structure of the modified films due to the annealing effect making a denser networking structure remained almost same even after further annealing at elevated temperatures. The minute change in the thickness is due to removal of hydrogen (table 3). Hence there are not significant different of thickness of the films when annealed at 750 and 1050 $^0$C.

## 4. Conclusions

We conclude that HDLC is composed of coherently coupled $sp^3$C-H and $sp^2$C=C carbons producing a continuous non-porous atomically smooth surface [5-6]. As prepared and annealed HDLC films, have been investigated by IBA techniques as described above by determining the depth profiling of hydrogen of the films and its thickness. This study reveals the nature of attachment of hydrogen atom to the carbon network of the film and also shows that the hydrogen content is maximum at the surface of the films. This hydrogen terminates the unsaturated bonds of carbon in the film surface at the end of their deposition. The hydrogen content increases with the increase of flow rate of methane at a constant flow rate of hydrogen and helium. The predominant character of graphitic layer towards the interface between the substrate and HDLC film have been substantiated from the thickness measurements by IAB techniques having high depth resolution of ~ 10 nm. The hydrogen content decreases with the depth of the films. This result matches with the Robertson model [25] of diamond/graphite/Si (100) interface which minimizes the lattice mismatch and interfacial energy. Therefore our HDLC film may be considered a graphene/graphitic structure, sandwich between the top diamond like structure ($sp^3$C-H) and silicon substrate. This material should have unique applications in future electronics. The studies described above, satisfactorily corroborate the expected variation in thicknesses of the film, both as prepared and annealed. These studies also exhibit the chemical phenomena like desorption of hydrogen from the film during both annealing process and bombardment with ion beams. Thus IAB techniques are very useful tools for the characterization of HDLC thin film. The denser networking structure, composed of carbon and hydrogen, remained almost the same even after the irradiation. This property should be useful for their application as protective coatings in the construction of radiation detectors.

**Acknowledgement**

NRR thanks the Department of Atomic Energy, Govt. of India, for financial assistance. We thank Mr. S. S. Sil and Mr. U. S. Sil, SINP for their technical help in this work.